\documentclass[%
reprint,
longbibliography,
amsmath,
amssymb,
aps,
prapplied,
floatfix,
twocolumn,
]{revtex4-1} 
\usepackage{graphicx}
\usepackage{siunitx}
\usepackage{amsmath}
\usepackage[symbol]{footmisc}
\renewcommand{\thefootnote}{\fnsymbol{footnote}}

\usepackage{hyperref}

\hypersetup{
    colorlinks=true,       
    linkcolor=blue,          
    citecolor=blue,        
    filecolor=blue,      
    urlcolor=blue           
}

\graphicspath{{./}{./Figures/}}

\begin{document}

\title{Confinement of an alkaline-earth element in a grating magneto-optical trap}

\author{A. Sitaram}
\author{P. K. Elgee}
\author{G. K. Campbell}
\email[]{gcampbe1@umd.edu}
\affiliation{Joint Quantum Institute, National Institute of Standards and Technology, Gaithersburg, MD 20899, USA}
\author{N. N. Klimov}
\author{S. Eckel}
\email[]{stephen.eckel@nist.gov}
\affiliation{Sensor Science Division, National Institute of Standards and Technology, Gaithersburg, MD 20899, USA}
\author{D. S. Barker}
\affiliation{Joint Quantum Institute, National Institute of Standards and Technology, Gaithersburg, MD 20899, USA}


\begin{abstract}

We demonstrate a compact magneto-optical trap (MOT) of alkaline-earth atoms using a nanofabricated diffraction grating chip.
A single input laser beam, resonant with the broad $^1$S$_0\,\rightarrow \,^1$P$_1$ transition of strontium, forms the MOT in combination with three diffracted beams from the grating chip and a magnetic field produced by permanent magnets.
A differential pumping tube limits the effect of the heated, effusive source on the background pressure in the trapping region.
The system has a total volume of around 2.4 L.
With our setup, we have trapped up to $5 \times 10^6$~$^{88}$Sr atoms, at a temperature of approximately $6$~mK, and with a trap lifetime of approximately 1~s.
Our results will aid the effort to miniaturize optical atomic clocks and other quantum technologies based on alkaline-earth atoms.
\end{abstract}



\maketitle

\section{Introduction}

Laser-cooled alkaline-earth atoms have applications in a wide range of quantum devices, including atomic clocks~\cite{Bothwell2019,Hinkley2013}, gravimeters~\cite{Sorrentino2010}, and spaceborne gravitational wave detectors~\cite{Graham2013, Vutha2015}.
The transition from a laboratory to field-based applications will require a drastic reduction in the size and complexity of laser-cooling systems.
For example, proposals to detect gravitational waves using alkaline-earth atoms require atom interferometers capable of being installed in satellites~\cite{Graham2013}.
Compact versions of these laser-cooled systems are also necessary in order make the unprecedented accuracy of alkaline-earth atomic clocks widely accessible~\cite{Parker2012}.

Laser-cooling experiments typically use a magneto-optical trap (MOT) to capture, cool, and confine the atoms.
Conventional MOTs use three orthogonal pairs of well-balanced, counterpropagating laser beams to confine atoms at the center of a quadrupole magnetic field.
As such, MOTs require large vacuum chambers with optical access along all axes, and have many degrees of freedom in alignment and polarization.
Compound optics can generate all necessary beams from a single input beam, reducing the complexity of the optical setup.
For example, pyramidal retro-reflectors maintain the beam geometry of conventional MOTs~\cite{Lee1996}.
However, the MOT forms inside the retro-reflecting optic, limiting optical access~\cite{Bowden2019, Pollock2009a}.
Tetrahedral reflectors form the MOT above the optic, maintaining optical access but breaking the geometry of a conventional trap by using only four beams~\cite{Vangeleyn2009}.
Tetrahedral MOTs can also be planarized by using diffraction gratings~\cite{Vangeleyn2010, Lee2013a, Nshii2013}.
Thus far, only experiments with alkali atoms have been successfully miniaturized using such grating MOTs~\cite{Barker2019, Lee2013a, Nshii2013}.
Here, we demonstrate a compact, grating MOT system for alkaline-earth atoms.

Alkaline-earth atoms pose unique challenges to miniaturization.
First, sources for alkaline-earth atoms must be heated to high temperatures (over 350~\si{\degreeCelsius}) to create sufficient flux of atoms to load a MOT.
Outgassing from the hot source can increase the background pressure, decreasing the trap lifetime, and equilibrium atom number.
Second, alkaline-earth atoms require large magnetic field gradients (on the order of 5~mT/cm), often created with large water cooled coils~\cite{Koller2017a}.
Third, with the high Doppler temperature of the broad $^1$S$_0\,\rightarrow\,^1$P$_1$ transition, and lack of sub-Doppler cooling, strontium and other alkaline-earth systems usually operate a second, subsequent MOT on the narrow $^1$S$_0\,\rightarrow\,^3$P$_1$ transition to achieve lower temperatures.
The ideal compact system must have the capability to operate at the two different cooling wavelengths.

\begin{figure}[t!]
  \center 
  \includegraphics[width=\columnwidth]{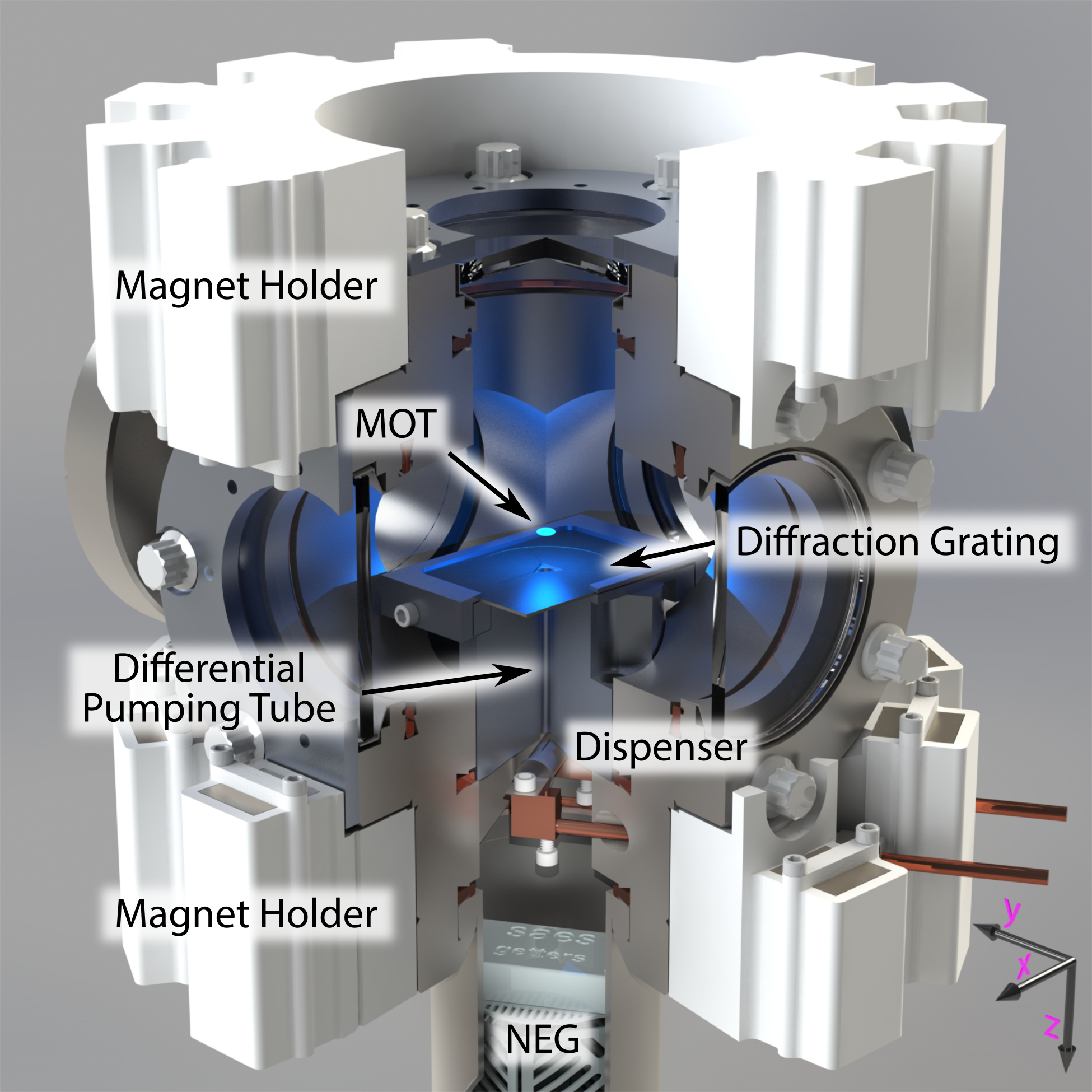}
  \caption{\label{fig:sketch}
  A cut-away model of the grating MOT system, with coordinates specified in the bottom right.
  The diffraction grating in the middle of the science chamber diffracts light to form the MOT beams. 
  The input laser beam (not shown) propagates along the $+\hat{z}$ direction.
  3D-printed magnet holders position permanent magnets around the top and bottom of the chamber.
  The dispenser source sits below the differential pumping tube and is pumped by a non-evaporable getter (NEG) pump.
  }
\end{figure}

Our system, designed around a diffraction grating chip, mitigates the above issues associated with miniaturizing a MOT for alkaline-earth atoms (see Fig.~\ref{fig:sketch}).
First, a 3~cm long differential pumping tube separates the vacuum chamber into two regions: the source chamber and the science chamber.
The source chamber contains a vacuum pump and a low-outgassing dispenser~\cite{Norrgard2018} that vaporizes strontium atoms.
The atoms then travel through the differential pumping tube before entering the science chamber.
Second, we create the magnetic field gradient for the MOT using permanent magnets, which are less complex than typical, water-cooled coils.
The magnetic field gradient extends into the differential pumping tube, forming an effective Zeeman slower when combined with the input laser beam.
Lastly, the first order diffraction efficiency of the grating we use is optimal at a wavelength of 600~nm, a middle ground between the two laser-cooling wavelengths (461~nm and 689~nm) for strontium.
Our compact alkaline-earth grating MOT system also maintains the optical access and achieves the atom number necessary for future quantum devices.

\section{Apparatus}

Our apparatus is shown in Fig.~\ref{fig:sketch}.
The vacuum system is comprised of two chambers, separated by a 3~cm long, 3~mm diameter differential pumping tube with an N$_2$ conductance of 0.11~L/s.
The MOT is located in a science chamber with four CF275 viewports, and pumped with a 75~L/s ion pump (not shown).
The source chamber is located below the differential pumping tube and is pumped with a 40~L/s non-evaporable getter (NEG) pump. 
Our source of Sr atoms is a 3D-printed titanium dispenser, described in Ref.~\cite{Norrgard2018}.
We run a current between 12~A and 14~A through the dispenser, effusing strontium towards the differential pumping tube.
Together, the source and science chambers are approximately 2.4~L in volume, although this estimate does not include the ion pump or the magnet holders.
The base vacuum pressure of $2\times10^{-7}$~Pa in the science chamber could be improved by replacing the ion pump with an NEG pump, which would also reduce the size of the apparatus.


The grating chip is located above the differential pumping tube, and has a triangular hole through its center, allowing atoms to enter the science chamber.
The grating chip was fabricated at the National Institute of Standards and Technology, and consists of three linear gratings arranged in a triangle.
The parameters of the chip are the same as those in Ref.~\cite{Barker2019}, except with a trench depth of \(150(2)~\si{\nano\meter}\).
This trench depth minimizes the $0^{\text{th}}$ order diffraction at 600~nm, which is between the 461~nm and 689~nm cooling transition wavelengths for strontium.
Each linear grating diffracts 32(1)~\% of the normally-incident 461~nm light into each of the $\pm1$ diffraction orders with an angle of \(27.0(5)\si{\degree}\).
For normally incident, circularly polarized light, the stokes parameters of the grating chip at \(461~\si{\nano\meter}\) are \(Q = -0.23(1),~U = -0.13(1),~V = 0.96(1)\), where $Q = 1$ ($Q=-1$) corresponds to $s$ ($p$) polarization defined relative to the plane of reflection for each linear grating.

Two sets of permanent magnets create the magnetic field for the MOT.
They are housed in 3D-printed magnet holders made of polylactic acid (PLA) that are designed to produce a compact setup with high magnetic field gradients, as shown in Fig~\ref{fig:sketch}.
Due to the geometric constraints of the vacuum chamber, the configuration of magnets is asymmetric, and the principal axes are rotated from those Fig~\ref{fig:sketch}.
We achieve maximum gradients of \{\(3.5~\si[per-mode=symbol]{\milli\tesla\per\centi\meter}\), \(2.7~\si[per-mode=symbol]{\milli\tesla\per\centi\meter}\), \(6.2~\si[per-mode=symbol]{\milli\tesla\per\centi\meter}\)\} along the \{$\hat{x}', \hat{y}', \hat{z}$\} axes, respectively, where $\hat{x}'$ and $\hat{y}'$ are rotated by $-\pi/6$ from $\hat{x}$ and $\hat{y}$.
By removing magnets from the holders, we can lower the gradient to 
 \{\(1.9~\si[per-mode=symbol]{\milli\tesla\per\centi\meter}\), \(1.9~\si[per-mode=symbol]{\milli\tesla\per\centi\meter}\), \(3.8~\si[per-mode=symbol]{\milli\tesla\per\centi\meter}\)\} along the \{$\hat{x}',\hat{y}', \hat{z}$\} axes, respectively. 
The field gradient extends to $z \approx 50$~mm, where $z = 0$ corresponds to the $\mathbf{B} = 0$ and $z \approx $ 40~mm corresponds to the position of the source.

A single laser beam, red-detuned from the $^1\mbox{S}_0 \rightarrow\,^1\mbox{P}_1$ transition at 461~nm, enters through the top viewport along the +$\hat{z}$ axis and is normally incident upon the diffraction grating chip.
The input MOT beam has a $1/e^2$ radius of 12~mm and a maximum power of 92 mW.
For the $^1$S$_0\,\rightarrow\,^1$P$_1$ transition with natural linewidth $\Gamma/2\pi=30.5$~MHz, $I_{\text{sat}} = 40.3$ mW/cm$^2$, giving a maximum peak $I/I_{\text{sat}} \approx 1$.
Intensities $I/I_{\text{sat}}$ reported herein always refer to the peak intensity of the input beam.
The central portion of the beam continues through the hole in the diffraction grating and through the differential pumping tube.
This beam, combined with the magnetic field gradient, allows for a small amount of initial slowing of the atoms, similar to a Zeeman slower.
Atoms can be lost from the MOT because the excited $^1$P$_1$ state decays at a rate of 610 s$^{-1}$ to the $^1$D$_2$ state, which in turn decays to the $^3$P manifold.
To mitigate the atom loss, two repump lasers, with wavelengths 679~nm and 707~nm, address the $^3$P$_0\,\rightarrow\,^3$S$_1$ and $^3$P$_2 \,\rightarrow\,^3$S$_1$ transitions, respectively.
More information on the repump scheme can be found in Ref.~\cite{Vogel1999}.
The repump beams are combined together on a 50/50 beam splitter, and then combined with the input MOT beam using a polarizing beam splitter.

We use absorption and fluorescence imaging along $\hat{x}$ to characterize the MOT.
Absorption images are taken after the MOT atom number equilibrates using a probe beam resonant with the $^1$S$_0$$\,\rightarrow\,$$^1$P$_1$ transition with $I/I_{\text{sat}} \approx 0.01$.
We use the atom number from the absorption images to calibrate the atom number extracted from fluorescence images taken during loading.
The Labscript suite software~\cite{Starkey2013} controls the experiment and data collection.
More detailed information on the laser systems can be found in Ref.~\cite{Pisenti2019}.

\section{Results}

\begin{figure}[t!]
  \center
  \includegraphics[width=\columnwidth]{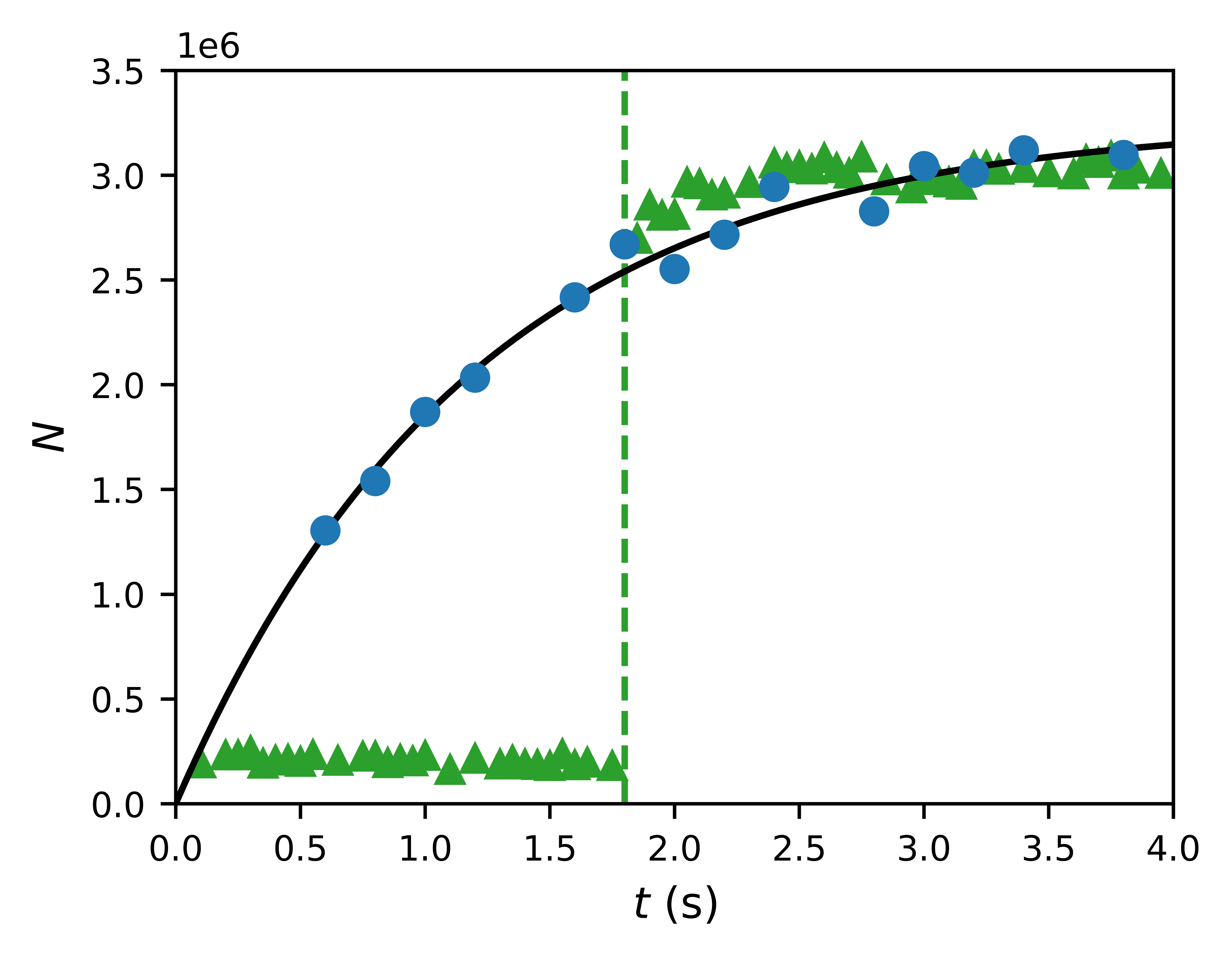}
  \caption{\label{fig:LoadingSequences} MOT loading curves with a source current of 13~A, $I/I_\text{sat}=1$, axial magnetic field gradient of $6.2~\si[per-mode=symbol]{\milli\tesla\per\centi\meter}$, and detuning $\Delta/\Gamma = -1$.
  The blue dots show the MOT atom number $N$ as a function of time $t$.
  The black curve is a fit to Eq.~\ref{eq:loading_curve}.
  The green triangles show the MOT loading without repump lasers, and subsequent recapture from the metastable reservoir.
  The dashed line indicates when the repump lasers were turned on.
}
\end{figure}

We measure atom number, loading rate, lifetime, and temperature to characterize the MOT.
During each experimental shot we take a sequence of fluorescence images while the MOT loads and construct a loading curve.
Fig.~\ref{fig:LoadingSequences} shows typical loading curves at an axial magnetic field gradient of $6.2~\si[per-mode=symbol]{\milli\tesla\per\centi\meter}$.
For a MOT with no light assisted collisions, the loading rate $R$, MOT lifetime $\tau$, and equilibrium atom number $N_0=R\tau$, are extracted by fitting each loading curve to the single exponential
\begin{equation}
    \label{eq:loading_curve}
    N(t) = R\tau(1-e^{-t/\tau}).
\end{equation}
An example fit is shown with a solid black curve in Fig.~\ref{fig:LoadingSequences}.
The quality of the fit to Eq.~(\ref{eq:loading_curve}) indicates light assisted collisions and secondary scattering are negligible.
At the higher gradient of 6.2~\si[per-mode=symbol]{\milli\tesla\per\centi\meter}, we observe typical loading rates of $4 \times 10^6$~s$^{-1}$ and a vacuum-limited lifetime of $1$~s.
We observe a similar loading curve at the lower gradient of $3.8$~mT/cm.

\begin{figure}
  \includegraphics[width = \columnwidth]{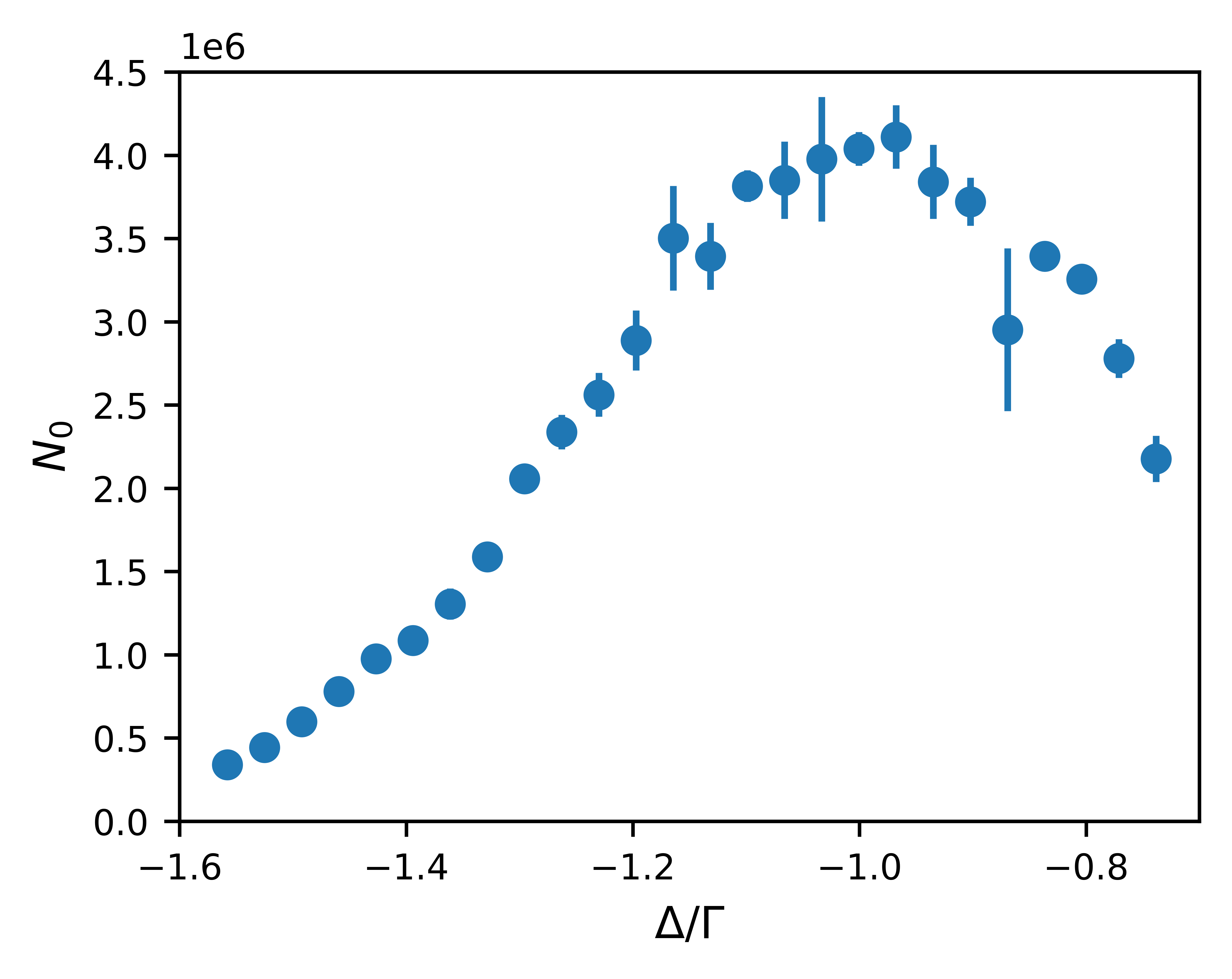}
  \caption{\label{fig:Detuning}The equilibrium atom number $N_0$ as a function of MOT beam detuning $\Delta/\Gamma$, with a source current of 13~A, axial magnetic field gradient of $6.2~\si[per-mode=symbol]{\milli\tesla\per\centi\meter}$, and $I/I_\text{sat}=1$.
  The optimal detuning is $\Delta/\Gamma = -1$.
  Most of the error bars are smaller than the data points.}
\end{figure}

\begin{figure}
  \includegraphics[width = \columnwidth]{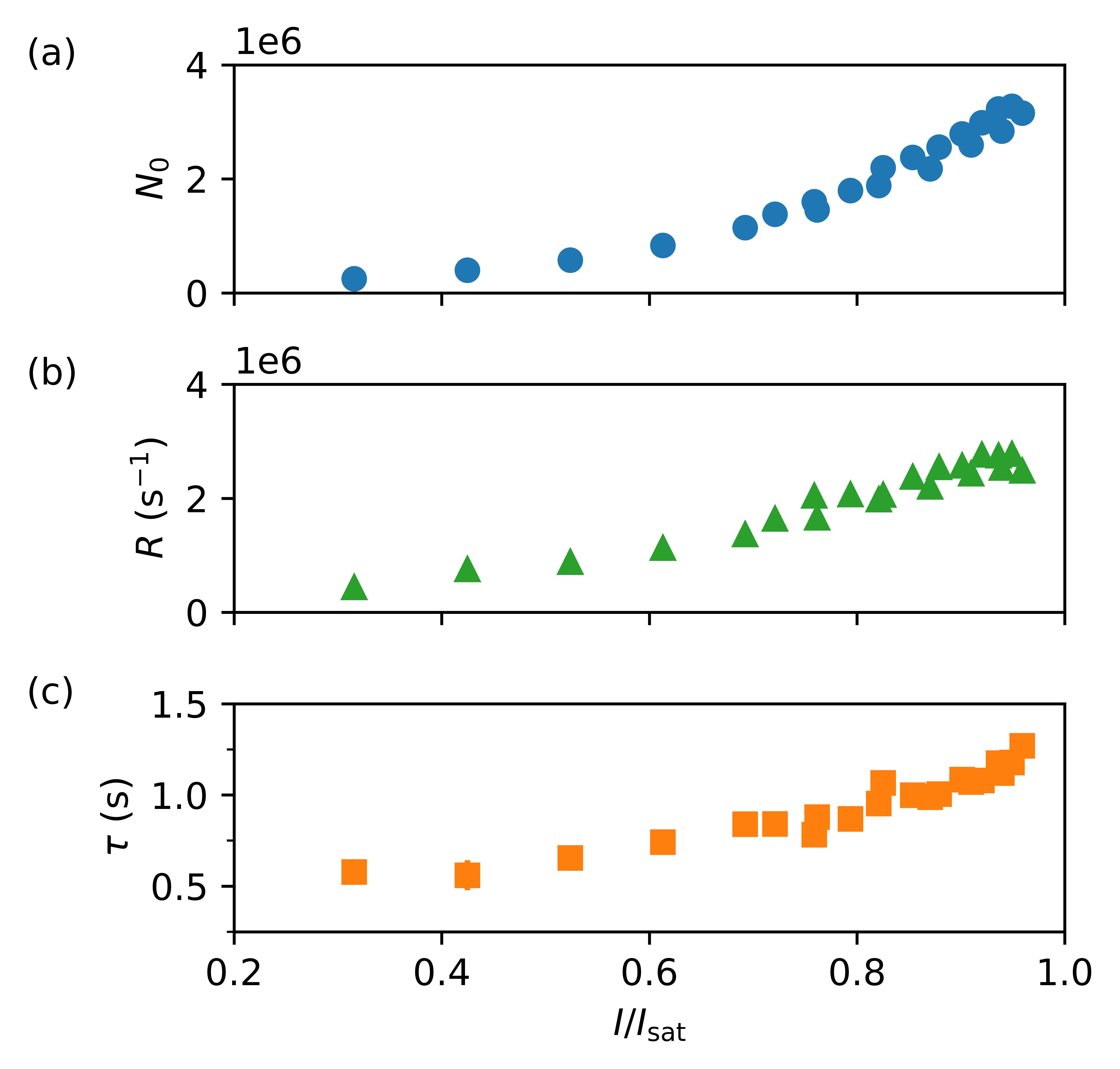}
  \caption{\label{fig:Intensity} MOT loading parameters as a function of $I/I_\text{sat}$: (a) equilibrium atom number $N_0$, (b) loading rate $R$, and (c) lifetime ($\tau$).
  Here, the source current is 13~A, the detuning $\Delta/\Gamma = -1$, and axial field gradient is $6.2~\si[per-mode=symbol]{\milli\tesla\per\centi\meter}$.
  The error bars on the points are comparable to the marker size.}
\end{figure}

Fig.~\ref{fig:Detuning} and Fig.~\ref{fig:Intensity} show the MOT parameters as a function of detuning from resonance and intensity, respectively.
We find the maximum atom number of approximately $4 \times 10^6$ at a source current of 13~A and $\Delta/\Gamma \approx -1$, a typical detuning for a conventional 6-beam Sr MOT~\cite{Xu2003, Schrek2009, Stellmer2013}.
As shown in Fig.~\ref{fig:Intensity}(a), the atom number continues to increase with $I/I_{\text{sat}}$, even at our maximum intensity, indicating that more laser power would be beneficial. 
The increase in $N_0=R\tau$ is only partially due to the increase in the loading rate $R$, shown in Fig.~\ref{fig:Intensity}(b).
Part of the atom number increase is due to an increase in the lifetime with intensity, shown in Fig.~\ref{fig:Intensity}(c).
The lifetime increase suggests that the trap depth is increasing with laser power, which in turn increases the escape velocity for a Sr atom that undergoes a background gas collision~\cite{Fagnan2009, Gensemer1997}.
However, the interplay between MOT temperature and tighter radial confinement with increasing intensity may also play a role.

We can also use the MOT to continuously load a magnetic trap, which consists of atoms that are trapped in the metastable $^3$P$_2$ state.
With strontium, the metastable magnetic trap is often used to increase the capture of rare isotopes and was key to the realization of quantum degeneracy~\cite{Schrek2009,Killian2009}.
By operating the MOT without repump light, atoms are shelved in the $^3$P$_2$ state where they are trapped by the MOT magnetic field.
When the repump light is turned on after atoms have accumulated in the magnetic trap, we see a sharp increase in the MOT atom number as shown in Fig.~\ref{fig:LoadingSequences}.
The recovery confirms that atoms are being caught and held in the magnetic trap, however we do not see a transient enhancement above the equilibrium atom number as demonstrated elsewhere~\cite{Nagel2003}.
Given our densities, and vacuum-limited atom number, we would not expect enhancement from magnetic trap loading.
Adding a depumping laser could enhance the loading rate of the magnetic trap and increase the atom number~\cite{Barker2015}.

\begin{figure}
  \center 
  \includegraphics[width=\columnwidth]{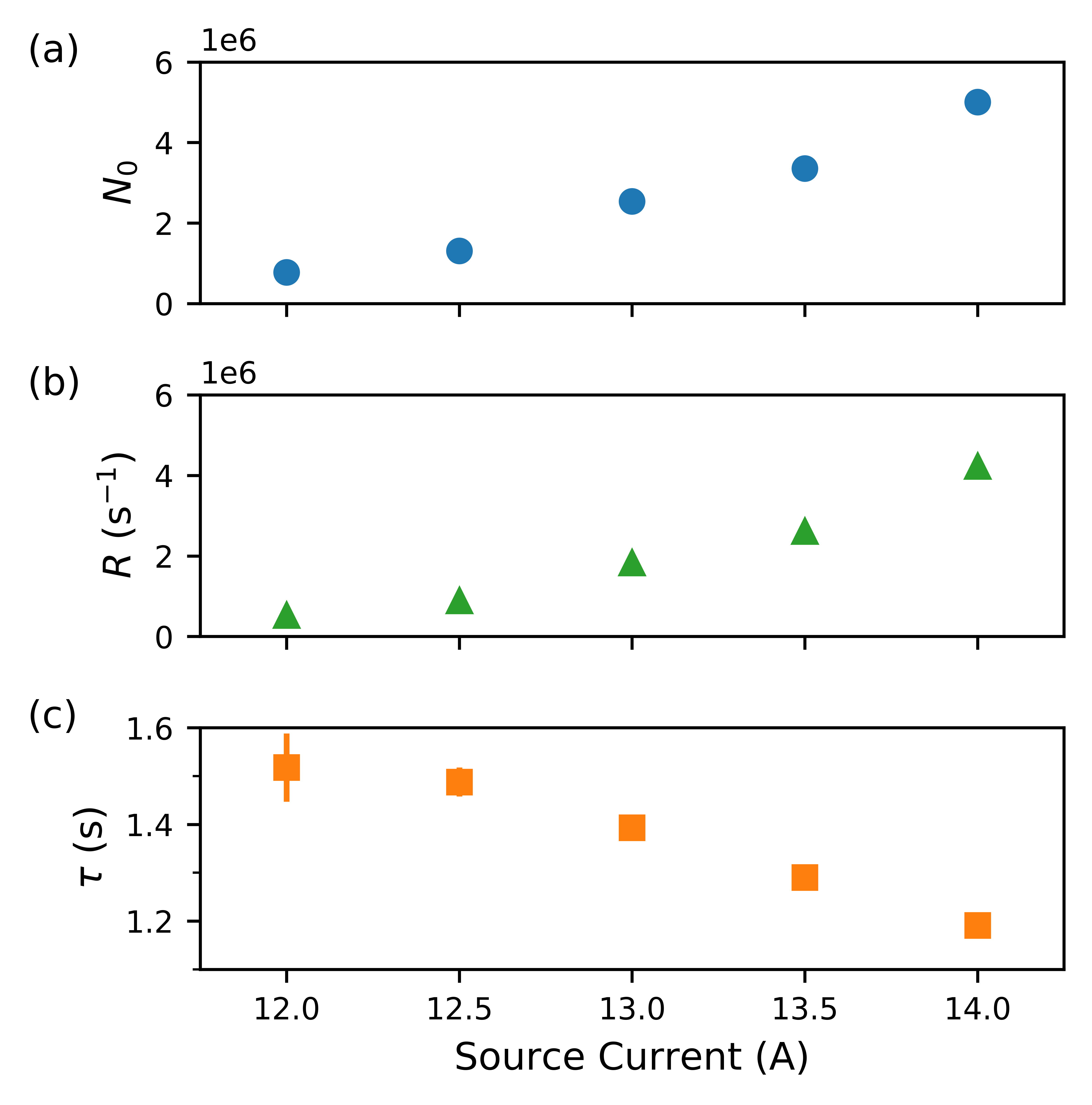}
  \caption{\label{fig:SourceCurrent}
    MOT loading parameters as a function of source current: (a) equilibrium atom number $N_0$, (b) loading rate $R$, and (c) lifetime ($\tau$).
    Here, $I/I_\text{sat} \approx 1$,  $\Delta/\Gamma = -1$, and axial field gradient is $6.2~\si[per-mode=symbol]{\milli\tesla\per\centi\meter}$.
    The error bars on most of the points are comparable to the marker size.}
\end{figure}

We investigate the effect of the source current on the atom number, loading rate, and lifetime, shown in Fig.~\ref{fig:SourceCurrent}.
The source current sets the temperature of the source, which in turn determines both the vapor pressure and the average velocity of atoms leaving the source.
At our highest achievable source current of $14$~A, limited by the ampacity of our electrical feedthroughs, we trap $5\times10^6$ atoms, but have still not saturated the atom number.
Based on the fit presented in Ref.~\cite{Norrgard2018}, we estimate the source temperature at 13~A to be over 600~\si{\degreeCelsius}.
When the source current is increased from 0~A to 13~A, the vacuum pressure in the science chamber increases by $3 \times 10^{-8}$~Pa, suggesting that the differential pumping is sufficient.
The increase in pressure is consistent with the small lifetime decrease shown in Fig.~\ref{fig:SourceCurrent}(c).

\begin{figure}
  \includegraphics[width = \columnwidth]{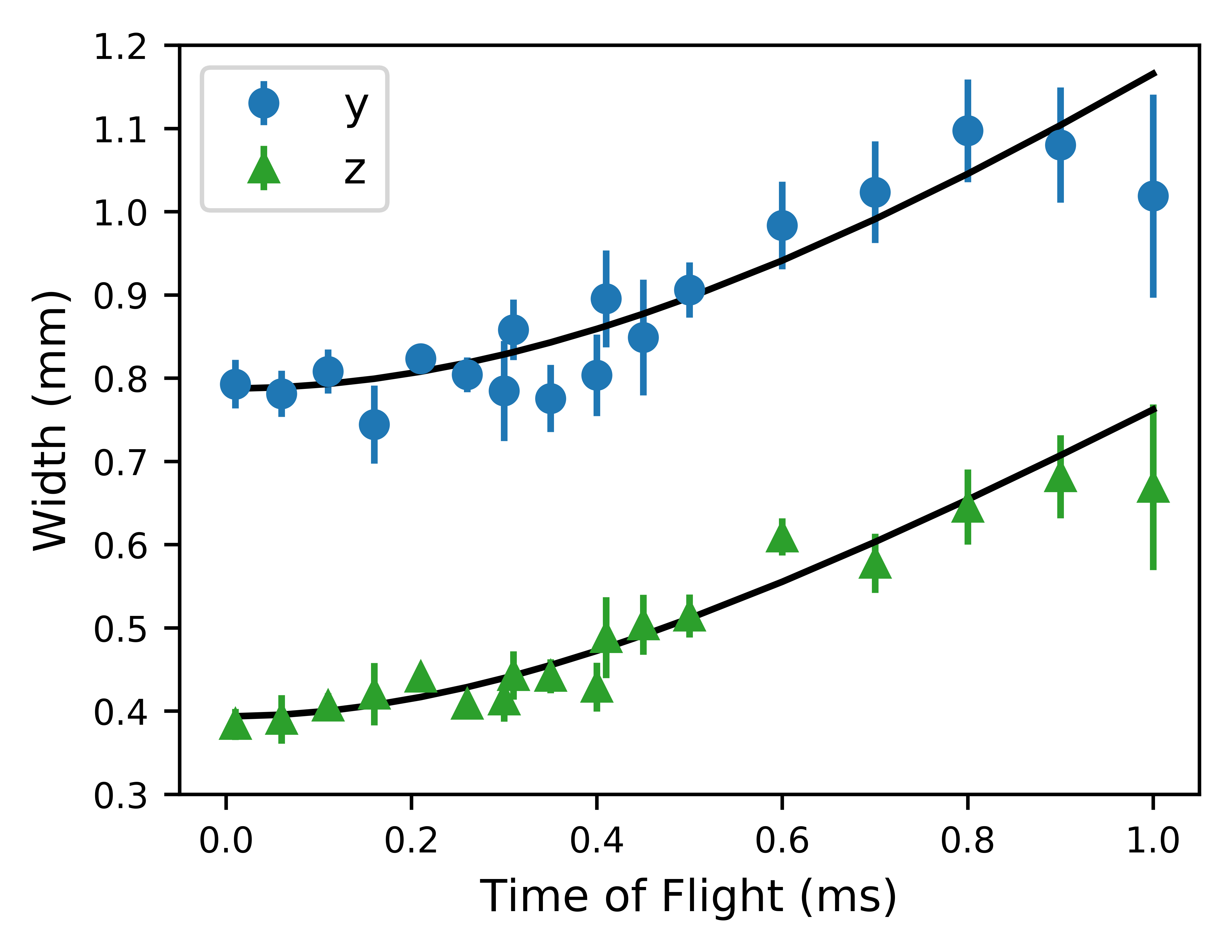}
  \caption{\label{fig:Temperature}Temperature measurement of the atomic cloud.
The rms width of the atomic cloud in the $\hat{y}$ direction (blue circles) and in the $\hat{z}$ direction (green triangles) are plotted against time of flight and fitted to the expansion function discussed in the text (black curves).
The calculated temperatures based on the fits are 7.8(9)~mK and 4.6(4)~mK for $\hat{y}$ and $\hat{z}$, respectively.
This data was taken with a source current of 13~A, $I/I_\text{sat}=1$, axial magnetic field gradient of $6.2~\si[per-mode=symbol]{\milli\tesla\per\centi\meter}$, and detuning $\Delta/\Gamma = -1$.}
\end{figure}

To determine the temperature of the MOT, we measure the width of the atomic cloud as it expands in time of flight, shown in Fig.~\ref{fig:Temperature}.
A Gaussian fit extracts the root-mean-square (rms) width, w, of the cloud in both the $\hat{y}$ and $\hat{z}$ directions.
The extracted widths are binned by time of flight, and the error bars are calculated from the standard error about the mean.
We fit the data to ${w(t)}^2 = w_{0}(t)^2 + v^2_{rms}t^2$, where $w_0$ is the initial rms width of the cloud, $v_{rms}=\sqrt{k_BT/m}$ is the rms velocity, $k_B$ is Boltzmann's constant, $m$ is the atomic mass, and $T$ is the temperature of the atomic cloud.
The temperature is 7.8(9)~mK and 4.6(4)~mK for ${\hat{y}}$ and ${\hat{z}}$, respectively, which is consistent with a conventional six-beam MOT~\cite{Xu2003}.
The temperature is not equal in the two dimensions because the diffusion coefficient and velocity damping constant are different in the axial and radial directions in a grating MOT.
For our MOT, the ratio of the temperatures along $\hat{y}$ and $\hat{z}$ is $1.7(2)$, consistent with the ratio of $1.9$ from the theory in Ref.~\cite{McGilligan2015}.

While the discussion has focused on trapping $^{88}$Sr, strontium has a number of stable isotopes. The isotope abundances for strontium are 82.58\%, 7.00\%, 9.86\%, and 0.56\% for $^{88}$Sr, $^{87}$Sr, $^{86}$Sr, and $^{84}$Sr, respectively.
Our setup can also trap around $7 \times 10^5$ atoms of $^{86}$Sr at a source current of 13~A, consistent with the abundances above.
Likewise, we would also expect to trap around $5 \times 10^5$ atoms of $^{87}$Sr, but were unable to realize a MOT of $^{87}$Sr.
The hyperfine structure of $^{87}$Sr poses at least two complications.
First, we might not have sufficient repump power to adequately address all necessary hyperfine transitions~\cite{Boyd2007}.
Second, the hyperfine structure combined with the non-trivial geometry and polarizations of the grating MOT may significantly weaken the confining forces~\cite{Mukaiyama2003, Lee2013a}.
The theoretical details of the latter are beyond the scope of this work and will be presented in a future publication.

\section{Discussion}

We have realized a grating MOT of alkaline-earth atoms in a compact 2.4~L apparatus.
Our permanent magnet design supplies the necessary field gradients for the MOT and allows for a degree of tunability, while the differential pumping tube limits outgassing from the hot source.
The MOT traps up to $5\times 10^6$ atoms of $^{88}$Sr at a loading rate of $4 \times 10^6$~s$^{-1}$, with a lifetime of approximately 1~s.
We also observe MOTs of $^{86}$Sr with $7\times10^5$ atoms, consistent with the relative isotopic abundance.



In the future, upgrades to our apparatus could be made to improve the performance of the MOT and decrease the size of the system.
We could improve the quality of the vacuum and reduce the size of the apparatus by replacing the ion pump with another NEG pump.
Improving the quality of the vacuum would increase atom number and lifetime, potentially allowing us to observe a MOT of $^{84}$Sr.
The system could be further miniaturized by using a fiber-coupled and photonically integrated chip to expand the MOT beam to the appropriate size without additional optics~\cite{Yulaev2019, Kim2018}.
By incorporating electromagnets with our permanent magnet assembly, we could adjust the magnetic field gradient to allow transfer of the atoms to a MOT operating on the narrow $^1$S$_0\,\rightarrow\,^3$P$_1$ transition.
The grating has good diffraction efficiency at both cooling wavelengths, and the diffracted beams have sufficient overlap to facilitate transfer between the MOTs.

The implementation of field-deployable quantum devices relies on compact systems.
Alkaline-earth-based quantum sensors have been proposed as platforms for atom interferometers and atomic clocks.
Compact interferometers could be used for inertial navigation~\cite{Hogan2008}, and gravitational wave detection in space~\cite{Graham2013}.
Deployable networks of optical clocks will be important for improved time and frequency metrology~\cite{Parker2012}, and tests of fundamental physics~\cite{Tino2007}.
Our results show that alkaline-earth grating MOTs are a promising step towards the development of compact optical clocks and other quantum devices.

\section*{Acknowledgements}

Ananya Sitaram and Peter Elgee contributed equally to this work.
We thank Francisco Salces Carcoba and Hector Sosa Martinez, for their careful reading of the manuscript.
We also thank the NIST Center for Nanoscale Science and Technology NanoFab staff for allowing us to use the facility to fabricate grating chips.
This work was partially supported by the NSF through the Physics Frontier Center at the Joint Quantum Institute.

\bibliography{sr_grating_MOT}

\end{document}